\documentclass[12pt]{article}
\usepackage{psfig}
\begin{document}

\newcommand{\lp}{\left}
\newcommand{\rp}{\right}
\newcommand{\be}{\begin{equation}}
\newcommand{\ee}{\end{equation}}

\title{Numerical study of the scaling properties of SU(2) 
lattice gauge theory in Palumbo non-compact regularization }
\author{
{Giuseppe Di Carlo\footnote{On leave from Laboratori Nazionali di
Frascati, INFN}} \\
{\small\it Laboratori Nazionali del Gran Sasso -- INFN} \\ 
{\small\it I-67010 Assergi (L'Aquila), Italy} \\ 
{\small e-mail address: {\tt giuseppe.dicarlo@lngs.infn.it}} \\
{Roberto Scimia} \\
{\small\it Dipartimento di Fisica and INFN -- Sezione di Perugia} \\ 
{\small\it Universit\`a degli Studi di Perugia} \\ 
{\small\it Via A. Pascoli I-06100 Perugia, Italy} \\ 
{\small e-mail address: {\tt roberto.scimia@lnf.infn.it}}
}
\maketitle
\begin{abstract}
In the framework of a non-compact lattice regularization of nonabelian 
gauge theories we look, in the $SU(2)$~case, for the scaling window through 
the analysis of the ratio of two 
masses of hadronic states. In the two-dimensional parameter space 
of the theory we find the region where the ratio is constant, and 
equal to the one in the Wilson regularization. In the scaling region 
we calculate the lattice spacing, finding it at least $20 \%$ larger 
than in the Wilson case; therefore the simulated physical volume is larger. 
\end{abstract}

\section{Introduction}

Lattice regularization is the most effective method, if not the only one, 
to study the behaviour of Quantum Field Theories 
outside the limits of application of perturbative expansions. 
One of the paradigms of this approach, namely the Wilson regularization
of gauge theories~\cite{Wil}, have been widely used from the beginning and
for a large period of time has been thought,
apart from few exploratory studies of alternatives, essentially 
as the only way, especially for non-abelian theories.
The Wilson regularization implies the use of gauge group elements as
dynamical (link) variables, instead of fields in the algebra of the 
gauge group as in the continuum. 
Therefore it is called a compact regularization, because the 
links take values in a compact space, the manifold of the
gauge group.
In the continuum limit one expects that the gauge fields
will pass through an effective decompactification restoring the properties
of the continuum physics.
Instead the naive discretization of gauge theories
using the usual field representation of the continuum formulation, 
replacing derivatives with finite differences and
a flat measure for the gauge fields, leads to a theory where gauge 
invariance is explicitely broken at finite lattice spacing.

The possible spurious effects of compactification have been
investigated in the past, leading to the conclusion that the main features
of non abelian gauge theories, e.g. confinement and spontaneous breaking of
chiral symmetry, do not depend on the compactification of dynamical
variables. Nevertheless one can wonder if there are alternatives
to this approach, and which are the possible advantages of a formulation
of a gauge theory on the lattice where the dynamical variables stay 
non-compact from the beginning.

Some schemes of non-compact regularization of lattice gauge theories have 
been proposed in the last ten years; we will concentrate our attention on
a particular one, the Palumbo regularization~\cite{Fab-PLB}, 
and we will determine in a
non-perturbative way the scaling region and the lattice spacing in physical
units. This regularization has been already studied and used in the past
both in the Lagrangian~\cite{BeFab1}\cite{BeFab2}\cite{FPV}, 
and Hamiltonian~\cite{Ted} formulation. Numerical results \cite {FPV} 
obtained in the Lagrangian formulation show a discrepancy with respect to
the perturbative expansion \cite{BeFab2}, while there is a full agreement
between the latter and the calculations in the Hamiltonian framework
\cite{Ted}; a reanalysis of this slightly controversial situation
is an additional (minor) motivation for the present work.

Besides being an alternative to 
Wilson regularization, Palumbo regularization is interesting because of 
its relation with the tadpole improvement tecnique used to obtain 
improvement of compact lattice actions  (see for 
example~\cite{Lep} and references therein). As will be shown in the 
following, in this regularization, because of the use of non-compact fields 
as dynamic variables, the tadpoles are resummed from the very beginning 
in some auxiliary fields which decouple in the continuum limit.

In the present study, in some sense an exploratory one, we will use
$SU(2)$ as gauge group, mainly for the significant simplification we get
in the numerical procedure
with respect to the (more interesting) $SU(3)$ case; but there are
no a priori obstructions in repeating the whole procedure we will depict
in the following for the $SU(3)$ case. Moreover in the $SU(2)$ case there
are, as said before, other results in the literature, useful for comparison.

We will proceed as follows: after the introduction of the more important 
features of the Palumbo regularization, we will explain our scheme for
identifying the scaling region using a ratio of two hadron masses.
Next step will be the definition of the scheme used in the numerical
simulations.
The results will be presented in the last section of the
paper, together with a comparision with other, analytical as well as
numerical, results obtained in the same framework in the past.

\section{The Palumbo non-compact regularization}

The non-compact regularization we used is fully explained 
in~\cite{BeFab1},\cite{BeFab2}, but in order to keep 
the paper self-contained we recall here 
its main features. The exact gauge invariance at finite lattice spacing 
is obtained by using a covariant derivative~${\cal{D}}_\mu (x)$ which, under 
a gauge transformation~$g\,\epsilon\, SU(N)$, transforms according to the 
equation 
\be
{\cal{D}}_\mu '(x) = g(x){\cal{D}}_\mu (x)g^{\dagger}(x+a\hat\mu),
\ee
where~${\cal{D}}$ is an element of~$GL(M,{\cal{C}})$, $M$~is 
the dimension of the matrix representation and $a$ is the lattice
spacing. The covariant derivative can 
then be written as a function of a field in the Lie algebra\footnote{As
shown in~\cite{BeFab1} it transforms like a continuum gauge field, except 
for lattice artifacts that vanish in the continuum limit.} of~$SU(N)$ plus 
some auxiliary fields, whose transformation equations can be worked out 
straightforwardly. In the case of the~$SU(2)$ gauge group, the covariant 
derivative in the fundamental representation depends on only one 
auxiliary field~$W_\mu$
\be
{\cal{D}}_\mu (x) = \lp (\frac{1}{a}-W_\mu\rp )I +i{\cal{A}}_\mu, \;\;\;\;
{\cal{A}}_\mu=A_\mu^a\,T_a,
\ee
where $I$ is the $(2\times 2)$ identity matrix and the~$T_a$ are the
generators of the gauge group
\be
\lp [T_a, T_b \rp ] = i\varepsilon^c_{ab}T_c,\;\;\;\; 
\lp \{T_a,T_b\rp \} =\frac{1}{2}\delta^a_b.
\ee
The strength tensor is defined in analogy to the continuum, 
and the same holds for the Yang-Mills lagrangian density
\be
{\cal{F}}_{\mu\nu}(x)=-i\lp [ {\cal{D}}_\mu (x){\cal{D}}_\nu (x+a\hat\mu)
-{\cal{D}}_\nu (x){\cal{D}}_\mu (x+a\hat\nu)\rp ],
\ee
\be
{\cal{L}}_{YM}(x)=\frac{\beta}{4}\sum_{\nu > \mu} \mbox{Tr}\,
{\cal{F}}_{\mu\nu}(x){\cal{F}}_{\mu\nu}^{\dagger}(x). \label{LYM}
\ee
The lattice theory is defined with a flat measure for 
the fields in~${\cal{D}}_\mu (x)$, as in the continuum. 

This is a regularization of the Yang-Mills theory 
if, in the continuum limit, the auxiliary field~$W_\mu$ is decoupled. This 
can be achieved at the quantum level by introducing a potential which gives 
a divergent mass to this field. The potential is constructed 
using the gauge invariant quantity~$t_\mu$
\be
t_\mu (x)I= {\cal{D}}_\mu (x){\cal{D}}_\mu^{\dagger} (x)-\frac{1}{a^2}I
=\lp [ \frac{1}{4}{\cal{A}}_\mu^2 (x)+W_\mu^2 (x)-\frac{2}{a}W_\mu\rp ]I,
\ee
therefore the basic non-compact lagrangian is obtained by adding 
to~${\cal{L}}_{YM}$ the potential
\be
{\cal{L}}_c (x)=\beta_c\sum_\mu t^2_\mu (x),\;\;\;\beta_c > 0 .
\ee
If in the continuum limit
\be
\beta_c \simeq \lp (a\Sigma\rp )^{2-\varepsilon},\;\;\;\varepsilon > 0,
\ee
where $\Sigma$ is a parameter with the dimension of a mass, the auxiliary 
field has a mass of the order of $a^{-\varepsilon / 2} 
\Sigma^{1-\varepsilon/2}$. As in~\cite{BeFab2} we confine
ourselves to the case~$\varepsilon=2$. 

The Wilson regularization can be obtained by eliminating the auxiliary 
field by imposing the constraint 
\be
t_\mu = 0. \label{t=0}
\ee
This condition produces a compactification of the covariant derivative 
which becomes
\be
{\cal{D}}_\mu=\frac{1}{a} U_\mu
\ee
where $U_\mu\,\epsilon\, SU(2)$ can be identified with the Wilson link 
variable. The imposition of the constraint~(\ref{t=0}) is equivalent to 
taking the 
limit~$\beta_c\rightarrow \infty$. This can be made explicitly by introducing 
a polar representation for the covariant derivative~\cite{BeFab2}; we notice 
that the Jacobian for the change of variables provides the Haar measure 
for the link variables~$U_\mu (x)$.

We stress that the coupling~$\beta_c$ is not an irrelevant one, because it is 
necessary to render the lattice theory a regularization of the Yang-Mills 
gauge theory. 

In~\cite{BeFab2} the properties of the regularization were studied using a 
perturbative approach and adopting a polar representation for the covariant 
derivative. 
We checked that the results obtained 
in numerical simulations do not depend (as it must be!) on the 
parametrization used for the covariant derivative. 
It is worth noticing that the use of a cartesian parametrization for the 
covariant
derivative makes evident one advantage of this non-compact regularization 
with respect to the compact ones: the number of vertices in 
perturbative calculations stays finite independently from the order 
considered.

Some irrelevant terms were introduced in the action, to make easier the 
perturbative analysis of the regularization. They are constructed from 
the gauge-invariant quantity~$t_\mu$, and cancel some contributions 
originated from~${\cal{L}}_{YM}$ in eq.~(\ref{LYM}) which depend only on 
the field~$t_\mu$, so that the auxiliary field does not propagate at 
tree level. By an appropriate choice of the 
couplings of the irrelevant terms the lattice theory can be defined in 
terms of only two parameters, $\beta,\gamma$ with
\be
\gamma^2=2\lp (\beta_c+\frac{3}{4} \beta\rp ).
\ee
The condition~$\beta_c > 0 $ correponds to 
\be
\gamma > \sqrt{\frac{3}{2}\beta}.
\label{gami}
\ee
To render possible a 
comparison with the previous perturbative~(see also~\cite{Ted}) and 
numerical~\cite{FPV} calculations we used the same lagrangian 
of~\cite{BeFab2}, explicitely
\be
{\cal{L}} = {\cal{L}}_{YM}+\frac{1}{8}\beta a^2 \sum_{\nu > \mu} 
\lp (\nabla_\mu t_\nu -\nabla_\nu t_\mu \rp )^2 
+\frac{1}{2}\gamma^2 \sum_\mu t_\mu^2
\label{act}
\ee
where
\be
\nabla_\mu f(x) = \frac{1}{a}\lp [ f(x+a\hat\mu)-f(x)\rp ].
\ee

\section{Non-perturbative determination of the scaling properties of the 
regularization}

The goal of our work is to answer to the question: {\it What about the 
convergence to continuum in this regularization?} 
In this form the question in misleading, because {\it removing the 
regularization} has different meanings depending on the context 
considered. In fact, in the one-coupling-constant lattice regularizations 
of Yang-Mills theory, like Wilson's one,  the evolution of the bare 
coupling $g^2=2N/\beta$ is obtained from the equation
\be
a\frac{d\,g(a)}{d\,a}=-\beta(g)=b_0 g^3(a)+b_1 g^5(a)+O(g^7) \label{betaeq}
\ee
where $b_0,b_1$ are the universal one and two loop coefficients of 
the beta function expansion. In a~$SU(N)$ pure gauge lattice theory they 
are 
\be
b_0=\frac{11}{3}\frac{N}{16\pi^2}\,;\,\,\,\,b_1=\frac{34}{3} 
\lp (\frac{N}{16\pi^2}\rp )^2. 
\ee
In perturbative calculations to take the continuum limit means to evaluate 
the limit for~$a\rightarrow 0$ of the renormalized quantities calculated.

In a numerical simulation the situation is very different, because the
calculated quantities are not in analytical form, moreover the lattice 
spacing is always finite. Therefore to take the continuum limit in this
case means to determine a region, in the parameters space of the lattice 
theory used, where the results (at finite lattice spacing) are as close 
as possible to their (experimental) continuum limit. Usually this region 
is characterized by a value as large as possible  of the correlation 
lenght of the system. 
There it is also possible to study the asymptotic scaling of the calculated 
quantities, i.e. in which measure they follow the 
perturbative scaling equations.
It is important to notice here that the size of the scaling violations 
depends on the quantity considered. As is well known, the ratio of two 
quantities with the same physical dimension (for example two masses) is 
constant (if not for scaling violations) in a scaling region 
which is usually larger than the asymptotic scaling window.
In presence of irrelevant couplings, the scaling of the physical quantities 
should not depend on their values.

In Palumbo non-compact regularization there are two coupling constants, 
as explained before, whose evolution as functions of the lattice spacing 
can be determined from equations analogous to eq.~(\ref{betaeq}). 
We stress that~$\gamma$ is not an irrelevant coupling (in the common 
meaning of the term), as explained above. In~\cite{BeFab2} the authors 
define the quantity~$\Lambda_{NC}$, that in the sequel we call 
{\em non-compact scale parameter}, using the expression
\be
\Lambda_{NC}^2 a^2 =\lp (1+\frac{b_1^2}{b_0^3} g^2 \rp)
\exp \lp\{-\frac{1}{b_0g^2}-\frac{b_1}{b_0^2}\ln \lp (b_0g^2\rp )\rp\}
\equiv F(g)
\label{alg}
\ee
which is exactly the same function obtained in the case of Wilson 
regularization, except for the fact that~$\Lambda_{NC}$ is a function 
of~$g^2=2N/\beta$ and~$\gamma$. 
Such a function can be evaluated in perturbation theory, 
solving the evolution equations for the two coupling constants.
Actually, in~\cite{BeFab2} the authors obtained
\be
\Lambda_{NC}=\Lambda_W \exp\lp\{-\frac{c}{g^2\gamma^2}\rp\} \label{plbg}
\ee 
where $\Lambda_W$ is the scale parameter for Wilson regularization and
\be
c=\frac{12}{11}\pi^2 \cdot 0.88323.
\ee
In the 
limit of vanishing lattice spacing we have~$g\rightarrow 0$, then 
eq.~(\ref{plbg}) would imply~$\Lambda_{NC}\rightarrow 0$, i.e. an 
inconsistency  of the regularization, if not for an appropriate evolution 
of~$\gamma$. On general grounds, given eq.~(\ref{plbg}) for~$\Lambda_{NC}$ 
as function of the bare couplings, to have~$\Lambda_{NC}\neq 0$ in the 
continuum limit it is necessary that
\be
\lim_{a\rightarrow 0} g^2 \gamma^2 =\kappa \neq 0.
\ee
Actually in~\cite{BeFab2} the authors obtained to one loop order
\be
\gamma =\frac{\gamma_1}{g^2}+\gamma_2
\ee
where $\gamma_1$ is an arbitrary constant and~$\gamma_2$ to be determined 
by an higher loop calculation, then in the limit~$a\rightarrow 0$ 
\be
\lim_{a\rightarrow 0} \Lambda_{NC} =\Lambda_W .
\ee
Such result means that in perturbative calculations with Palumbo  
regularization it is possible to get the continuum limit with a scale 
parameter equal to Wilson one. 

In numerical simulations the situation is different, as explained above. 
On general grounds we expect that there will exist a scaling region in the 
plane~$(\beta,1/\gamma)\;$\footnote{We use as natural variable~$1/\gamma$ 
because the Wilson limit can be identified with the $1/\gamma=0$ line.}
 and that, because of the finite lattice spacing, 
the properties of convergence to the continuum of the regularization vary in 
this region. For example, we could have, in scaling conditions, different 
physical values for the lattice spacing (therefore different physical 
volumes) as a function of the value of the bare 
parameters~$\beta,1/\gamma$.
One possible strategy to determine the scaling properties of the 
regularization (which we followed) goes through the following steps. 
The ratio of two quantities with the same physical 
dimension has to be calculated  as a function of the bare parameters 
of the regularization, for example on a regular grid 
in the~$(\beta,1/\gamma)$ plane.  
These values can be fitted so as to obtain a 
surface continuosly varying in the~$(\beta,1/\gamma)$ plane. 
The scaling region for the non-compact regularization 
is therefore the region in the~$(\beta,1/\gamma)$ plane where the ratio 
considered agrees with the same quantity evaluated in Wilson lattice 
theory, which corresponds to the~$1/\gamma=0$ line. The comparison with
Wilson regularization (or with any other regularization for which 
perturbative relations are avaliable to determine the asymptotic scaling 
region) is mandatory only in the case of a non-physical theory, like in 
pure~$SU(2)$ gauge theory. In fact in the~$SU(3)$ case we can fix the 
physical value of the ratio by using experimental data, therefore in a 
fully independent way from perturbative calculations.

In summary, the only two necessary ingredients are the ratio calculated 
using the non-compact regularization, and the perturbative scaling 
equations for Wilson regularization to determine the physical value of 
the ratio, which is the value it assumes in the asymptotic scaling region 
for Wilson regularization. We stress that this tecnique was well known in 
principle (see for example~\cite{Collins}) and has been used in the past, 
altough in a different form, for example in~\cite{GDC}. 
We choosed to use the ratio of two particle masses, being led to this 
choice by the following considerations: other pure gluonic observables, 
like the plaquette, have significant perturbative contributions which may 
obscure the non-perturbative features of the regularization. As for the
glueball masses, usually smearing tecniques are used to obtain clearer 
signals, but this imply the introduction of additional parameters besides
the two peculiar to the Palumbo regularization. As for the string tension, 
in this preliminary work we preferred to use ratios of quantities with the 
same physical dimension to avoid spurious effects. Lastly we notice that 
the particle masses have a well defined
physical meaning and depend in a fundamental way on the non-perturbative 
properties of the theory.

It is a crucial condition for the above depicted scheme to be
valid, that the only dimensionful quantity of the theory is the 
renormalization group scaling parameter; this is true if we work in the chiral
limit, otherwise we would have another dimensionful quantity 
( the quark mass ).
As is well known it is extremely difficult (and numerically very 
expensive) to perfom simulation or to measure masses directly 
in the chiral limit; to stay within the limits of this work we have chosen
to evaluate the mass spectrum at four finite quark mass values and then
extrapolate to the chiral limit. To have a better control on this extrapolation
we have used Kogut-Susskind fermions, where chiral limit can be
easily defined and be reached, with a good accuracy, by means of
a linear extrapolation in the bare quark mass.
The same considerations led us to work in the quenched approximation;
moreover one of our (minor) scopes is to compare with the analytical 
calculations in~\cite{BeFab2} that were carried out for the pure 
gauge theory.

\section{Details on the numerical approach}

As follows from the previous considerations the action for SU(2) lattice gauge
theory using the Palumbo regularization contains two parameter ( $\beta$ and
$\gamma$ ); the role of $\gamma$ is to assure the decoupling of the 
auxiliary field in the continuum limit. 
Notice that, in the practice, the action can assume different 
specific forms, depending
on which class of irrelevant terms we decide to include; different choices
of the set of irrelevant terms can lead to an action easier to use in
numerical or analytical calculations, or to an action more complex, but 
with a better approach to the continuum limit. We will not address this
issue: we choose to work with the lattice action in~eq.~(\ref{act}).

We have explicitly checked that the term
\be
\frac{1}{8}\beta a^2 \sum_{\nu > \mu} 
\lp (\nabla_\mu t_\nu -\nabla_\nu t_\mu \rp )^2 
\ee
is actually irrelevant, in the sense that the results for the hadron
masses do not depend in a sensible way on its inclusion or exclusion
in the action.

The action (\ref{act}) can usefully be thought of as 
an action for a gauge field 
living in the GL(2) group, therefore we decided to use a cartesian 
representation that includes both the physical fields and the auxiliary one.
Other choices are possible; in particular we recall the polar representation
used in~\cite{BeFab2}.
Starting from the action~(\ref{act})
we have written a generic Metropolis+Overrelaxation code; the overrelaxation
part of the procedure applies only to the SU(2) part of the GL(2) fields,
i.e. it amounts to a microcanonical rotation in the SU(2) subgroup leaving
untouched the determinant of the GL(2) matrix.
CPU time and memory requirement for the computation is essentially 
the same needed for a corresponding MonteCarlo with the Wilson regularization.

Looking at the action~(\ref{act}) it can be easily understood that the 
parameter $\gamma$ is bounded to be larger than a minimum value 
(eq.~\ref{gami}) as discussed in~\cite{BeFab2}.

The other limit, i.e. $\gamma \to \infty$, reproduces the Wilson 
regularization, in the sense that the determinant of GL(2) gauge fields 
is constrained to be one and we recover the usual compact SU(2) gauge action.

In order to have a better readability of the results, in particular 
in the region of large $\gamma$ where the Wilson results have to be
recovered, we decided to work on a (quasi) regular grid in the
$\beta$,$1/\gamma$ parameter space (in this space the $1/\gamma =0$ line
is the Wilson theory). We have chosen to work with $2.0\leq\beta\leq 2.7$ 
and $0.0<1/\gamma\leq 0.2$, using 12 values of $\beta$ and 20 values 
of $1/\gamma$; we have also included in our analysis the results 
obtained for the Wilson regularization (the $1/\gamma=0$ line in
the following).

As said before we are interested mainly in the masses of the (lighter)
hadron states;
in $SU(2)$ gauge theory we have 4 states made from 2 quarks,
namely a scalar ($\sigma$),
a pseudoscalar ($\pi$), a vector ($\rho$) and a pseudovector ($A_1$).
Moreover, due to the use of Kogut-Susskind fermions, we can have a
signal in the correlation function both from non-oscillating and
oscillating channel; we will refer to them as $+$ and $-$ state, 
respectively. 

We would like to stress here that we had $240$ different simulations to
 carry out in order to complete our program; facing with our limited
computing resources, and taking into account the exploratory character
of this work, we confined ourselves to small lattices, namely a
$6^3\times 12$ one. 

The actual scheme we have followed is: for every point in the 
$\beta,1/\gamma$ grid we have thermalised a starting configuration, then 
measured the hadron propagator for one of the four values of the quark 
mass (namely $m_q=0.15,\;0.20,\;0.25,\;0.30$), on a configuration separated 
by $150$ cicles of combined Metropolis+overrelaxation sweeps from the 
previous one used to measure the observables, for a total of $180$ 
hadron propagators per mass value.

We notice that we have 
estimated the integrated autocorrelation time in a representative 
point inside the scaling window, finding a value around $40$,
therefore we used an ensemble of well decorrelated configurations. 

From the averaged propagators we have extracted the mass of hadron
states ($m^+$ and $m^-$) fitting with the following function:
\be
P(\tau)=A \lp ( e^{-\tau m^+}+e^{-\lp (N_t-\tau\rp ) m^+}\rp )+
B \lp (-1\rp )^\tau \lp ( e^{-\tau m^-}+e^{-\lp (N_t-\tau\rp ) m^-}\rp )
\label{fitpro}
\ee
At the end we obtain, for each $\beta,1/\gamma$ point, the mass of
seven hadron states (for the pseudoscalar channel the oscillating
state is not observable) evaluated each one at four values of the
quark mass. 

Among these seven masses
we notice a well defined pattern. The pion mass is affected by the
smallest statistical error (well below $1\%$), but, this particle beeing 
a Goldstone boson, it vanishes in the chiral limit, and then can not be 
used for the determination of the scaling window.
Other three particles, namely the $\rho^+$, the $A_1^-$ and the
$\sigma^+$, have small statistical errors (around or less than $1\%$);
finally the other three states are worst defined beeing affected by
large statistical errors, and then useless for our purposes.

Restricting to the three non-Goldstone states with small errors,
a linear extrapolation to the chiral ($m_q\to 0$) limit gives 
us the dataset for the analysis explained in the following section. 

Aside from the determination of the hadron propagators, we have 
used the configuration generated also to measure local
observables as the plaquette and its specific heat.

The simulations have been fully performed on small systems like
Unix workstations and Linux PCs at L.N.F. , L.N.G.S. and University
of Perugia.

\section{Results}

Let us proceed to show our results; as said in the previous section we have
evaluated the ratio of the masses of $A_1^-$ and $\rho^+$. The motivation
for this choice comes from the observation that the errors 
for these two masses are smaller; nevertheless we have
repeated the whole analysis also using the mass of $\sigma^+$ particle,
finding similar results.

The mass ratio $R$ has been obtained on a grid of points in $\beta$ and
$1/\gamma$ and then a regular surface has been reconstructed using a
bipolynomial spline fitting procedure. We postpone a discussion on the 
errors to the end of this Section.

In Fig. 1 we report the fitting surface $R(\beta,1/\gamma)$.
Looking at the figure we can recognise some important features;
first of all the existence of a (almost) flat region ({\it valley})
that originates
from the $1/\gamma=0$ line (Wilson results), and propagates towards 
larger values of $\beta$ for increasing $1/\gamma$. The flat region in 
the Wilson limit coincides with the usual scaling region for
$SU(2)$ pure gauge lattice theory for these (intermediate) lattice 
sizes~\cite{wsca}.

We have checked that, in this region, the good asymptotic scaling can
be obtained only in a narrow interval near $\beta=2.3$. 
Therefore we tentatively identify the valley as the scaling
window (although not the asymptotic scaling region) for the non-compact
regularization.
In order to make this observation more precise we report in Fig. 2 the curves 
of constant $R$ in the plane $\beta,1/\gamma$. 

Including in the analysis the data of the single particle masses in
the scaling region we can make our statement more definitive.
Looking at Fig.2 we can identify two different zones where a more 
detailed analysis allows us to clarify
the properties of the region we proposed as the scaling region.
If we define the scaling region as limited from the
$R=1.09$ level, we can say that $1/\gamma<0.12$ and $0.10<1/\gamma<0.17$ 
are the scaling regions for, respectively, $\beta=2.35$ and $\beta=2.60$.
Consider the lines $\beta=2.35$ and $\beta=2.60$: in the first case the
Wilson point ($1/\gamma=0$) is inside the scaling region, whereas in the
second case only a segment $1/\gamma_1<1/\gamma<1/\gamma_2$ does.
In Fig. 3 we can see the behaviour of $R$ (from the fitting surface)
along these two lines.

In Fig. 4-a,b we report the results for the $\rho^+$ mass in this two
cases ($\beta=2.35$ for Fig.4-a and $\beta=2.60$ for Fig.4-b). In these
figures we have reported the raw data for the mass.

We are allowed, now, to compare our numerical results 
with the perturbative analysis 
in~\cite{BeFab2}; following this analysis we expect that the lattice
spacing, and then the lattice mass inside the scaling region, follows
a behaviour like:
\be 
a(\beta,\gamma)\propto\exp{\lp\{-\frac{12\pi^2}{11} 0.2208 
\frac{\beta}{\gamma^2}\rp\}}
\ee
where the numerical coefficients result from an one loop calculation.

Coming back to Figs. 4 we can see, superimposed to the data in the
scaling regions, a fit with the exponential of a second order
polynomial. Notice that in the $\beta=2.35$ case
we have not included the Wilson point in the fitted data; the Wilson result 
is in good agreement with the extrapolation from $1/\gamma>0$ data; in
the other case we expect this kind of extrapolation to be meaningless.
For $\beta=2.35$ we can try a direct comparision with the perturbative 
results, keeping in mind that we are working on a small lattice and, in
any case, we have not performed a detailed estimation of systematic
effects, largely outside the scope of this work. 
Following~\cite{BeFab2} we expect the coefficient of the linear term
in $1/\gamma$ to be zero and that of the quadratic one to be 5.56.
From our fit we get for the former a value compatible with zero 
($0.3\pm 0.8$) and for the latter $7\pm 2$.
We do not claim an agreement with the perturbative formula, but
in any case we have, at this level, no trace of the large discrepancies
found in~\cite{FPV}.

Finally, in Fig. 4-b we note that the decreasing trend for decreasing
$1/\gamma$ which can be clearly seen in the scaling region ceases
in correspondence of the lower limit of the scaling region itself.
We have checked that this behaviour is present also in the data for
the other particles and for other values of $\beta$ where, as in the case
of $\beta=2.6$, the Wilson limit lies outside the scaling window.
Again, the behaviour of the mass, and hence the lattice spacing, is
well reproduced by an exponential of a polynomial in $1/\gamma$. 

Fortified by this results we can proceed, now, to a final check on
our scaling window; we expect that the scaling window contains the
value of the parameters for which the specific heat has a maximum,
signalling a large correlation lenght.
In Fig. 5 we can see the constant $R$ lines as in Fig. 2 (continuous lines) 
with superimposed the position, in the $(\beta,1/\gamma)$ plane, of the 
peaks of the plaquette specific heat (dashed line). In this figure four 
lines of constant plaquette (dotted lines) are also
reported.
We can see a substantial correspondance between the fluxes in the
parameter space as identified by different operators.
This scenario is the one expected for a honest theory in scaling regime.
We therefore are confident to have correctly identified 
the scaling window for non-compact regularization.
We stress that using this scheme, we can leave aside any perturbative 
calculation in the non-compact regularization.

Any point inside the scaling window is a good one for approximating
the behaviour of continuum theory, but actual results can be different. 
In particular the value of the lattice spacing, and then the lattice 
volume, varies from point to point. 
As we have seen before for the non-compact formulation this 
value can be larger than that obtained using the Wilson formulation.
In order to make this assertion less qualitative we present
in Fig. 6 the behaviour of the lattice spacing, as extracted
from the $\rho^+$ mass, along the {\it center} of the valley.
We can clearly see that the lattice spacing becomes larger and larger 
the more we depart from the Wilson case $1/\gamma=0$; with the lattice
size used in this work we can access to regions where the lattice 
spacing is around $20\%$ larger than the Wilson regularization one.
This improvement can be made larger if we move towards
larger $\beta$ and $1/\gamma$, but with a narrower scaling valley
and nearer to the instability regime (see eq.\ref{gami}). 
Again this results are in substantial agreement with the prediction
of the perturbative calculation in \cite{BeFab2},\cite{Ted} and do not show 
any sign of the large deviations claimed in \cite{FPV}; we remind, however,
that these deviations have been obtained using a different approach and
looking at different operators.

We believe that a complete knowledge of the entire scaling region, as well as
the corresponding gain in terms of lattice spacing, is out of the 
scope of this work; it is more interesting to address this point in a
more realistic simulation, using larger lattices in the $SU(3)$ case.
In the present paper we have devoted our attention more to the development
of a scheme for addressing the problem, than to give definitive, and 
quantitative, answers to questions about the (improving) potentiality of
this regularization.

\subsection*{ Errors}

The main ingredient of our analysis is the ratio $R$. This quantity
is obtained, starting from the raw propagators, by means of a
complex procedure, amounting essentially to three levels of fitting:
fitting the correlator to extract the mass, fitting the masses to
extract the chiral limit and eventually the final fit of $R(\beta,1/\gamma)$
to get a smooth surface. It is extremely hard
to trace the propagation of statistical errors from the raw data 
to the final surface.
Nevertheless it is mandatory to have at least a rough idea of the 
effects of the statistical fluctuations on our procedure. 

To this end we have chosen to proceed in this way: we have divided our 
statistical
ensemble of $180$ independent measures of the correlator (for each
$\beta$, $1/\gamma$ and quark mass) in two independent subsets of $90$; 
we have then repeated the whole procedure (from the fitting to 
eq.~\ref{fitpro} to the construction of the smooth surface for $R$) 
for the two sets independently. We have computed, then, the root mean 
square of the deviation between the two surfaces on a regular grid of
$O(100)$ points $P_i$, placed in the core of the $(\beta,1/\gamma)$ region
to avoid edge effects. 
From this procedure we find a r.m.s. of $1\%$ which we assume to
approximate the error on the $R$ surface;
we have checked the independence of 
this evaluation on the number of points $P_i$ used to compute the average
of the deviation. This result helped us to establish the criterion
($R\leq 1.09$) used to define the scaling region (see Fig.3)
and made us more confident on the robustness of the emerging scenario.

\section{ Conclusions}

We have studied the approach to the continuum of SU(2) lattice gauge theory 
within a non-compact regularization scheme with two dimensionless parameters.
We have determined the scaling window using a non perturbative approach,
defined through the use of the ratio of the masses of two hadronic
states. We have found a clear scaling window that, stemming from the
one of the Wilson regularization, moves towards larger $\beta$ and 
$1/\gamma$. 

Inside this region we have determined the lattice spacing, finding it
increasing with increasing distance from the Wilson line (in the
parameter space) and in a way compatible with the expectations based
on perturbative analytical calculations, without the large discrepancies
observed in~\cite{FPV}. 
Our determination of the scaling region is corroborated by the
observation of the behaviour of other quantities, like the specific 
heat, linked to the correlation length.

All this work has been carried out in small lattices and then we can
not give a more sound quantitative estimation of the effects observed.
In any case the use of this non-compact regularization leads to
clear advantages in term of simulated physical volumes.

This analysis needs to be improved with the use of
larger lattices and other observables more suitable for the 
accurate determination of the lattice spacing, possibly
for the physically interesting case of SU(3) theory.
It would also be of interest to study the effects of unquenching on the 
advantages for the physical volume in the scaling region in Palumbo 
regularization.

{\bf Acknowledgements}

We thank F. Palumbo and S. Caracciolo for invaluables discussions and 
suggestions during every stages of this work.
We thank M.P. Lombardo for providing us the program for Dirac matrix
inversion and propagator calculation.

\newpage

{\bf Figure Captions}

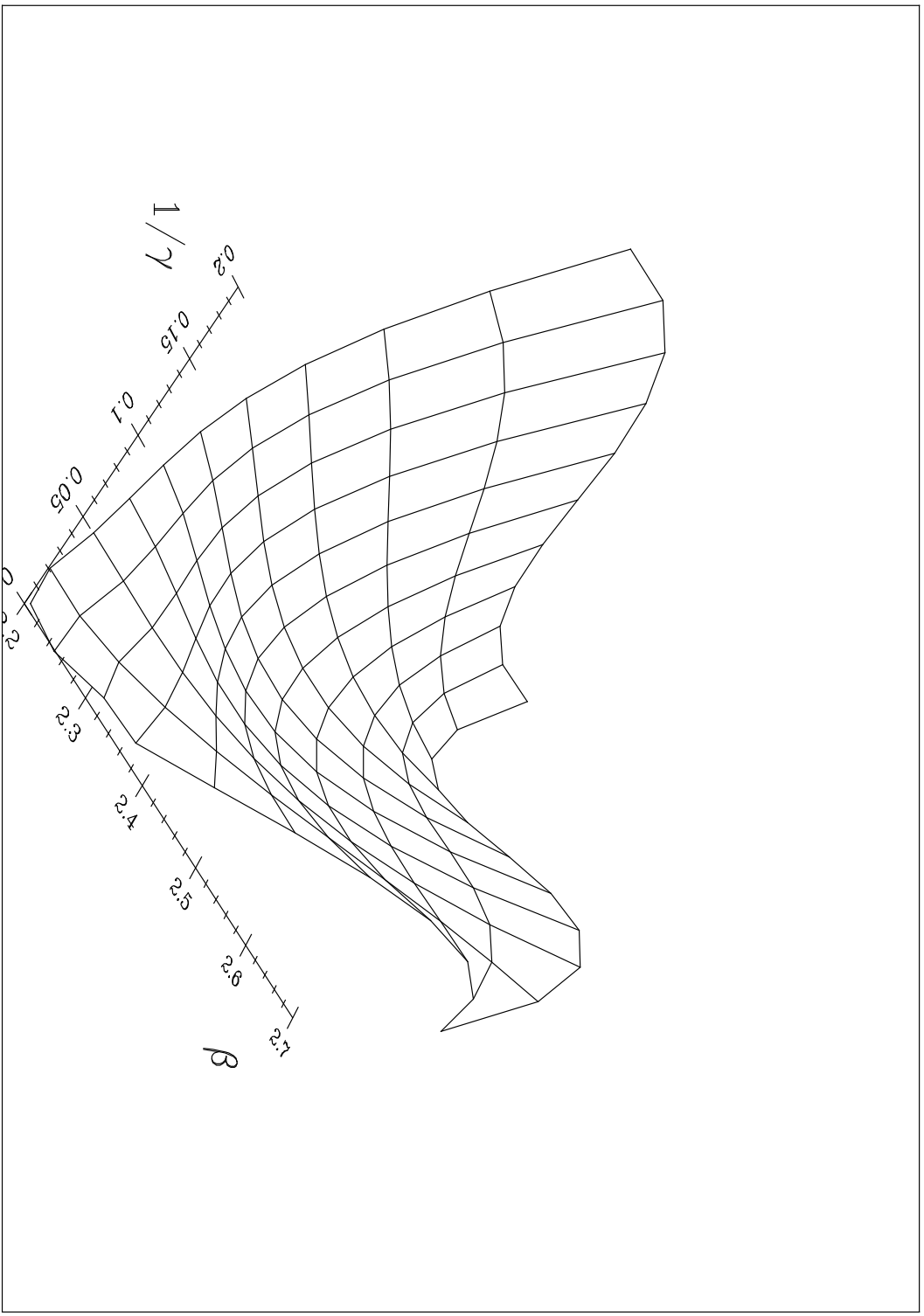
Fig. 1: $R(\beta,1/\gamma)$ fitting surface.
\vskip 0.35 truecm

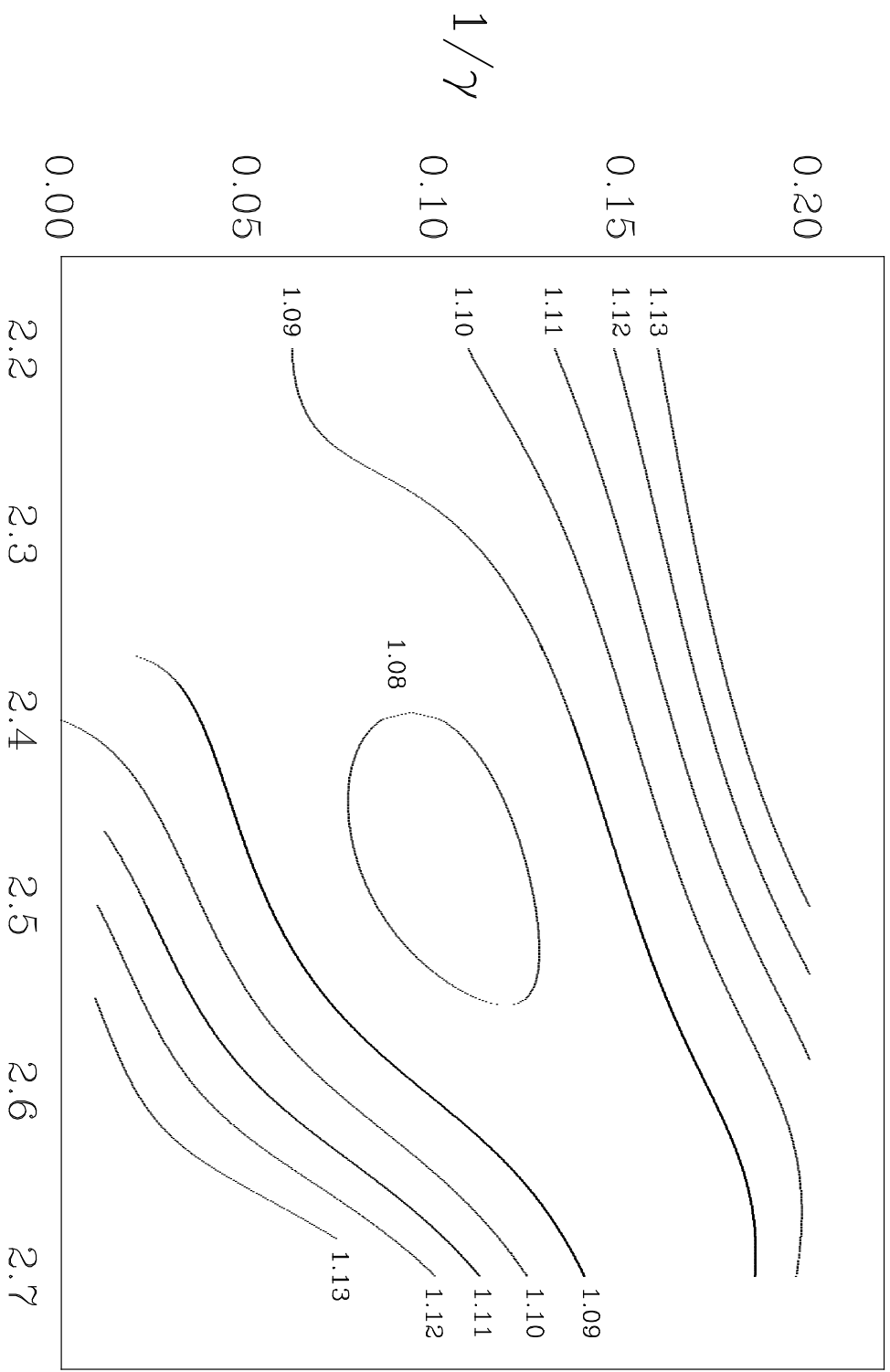
Fig. 2: Lines of constant $R$ in the $\beta,1/\gamma$ plane.
\vskip 0.35 truecm

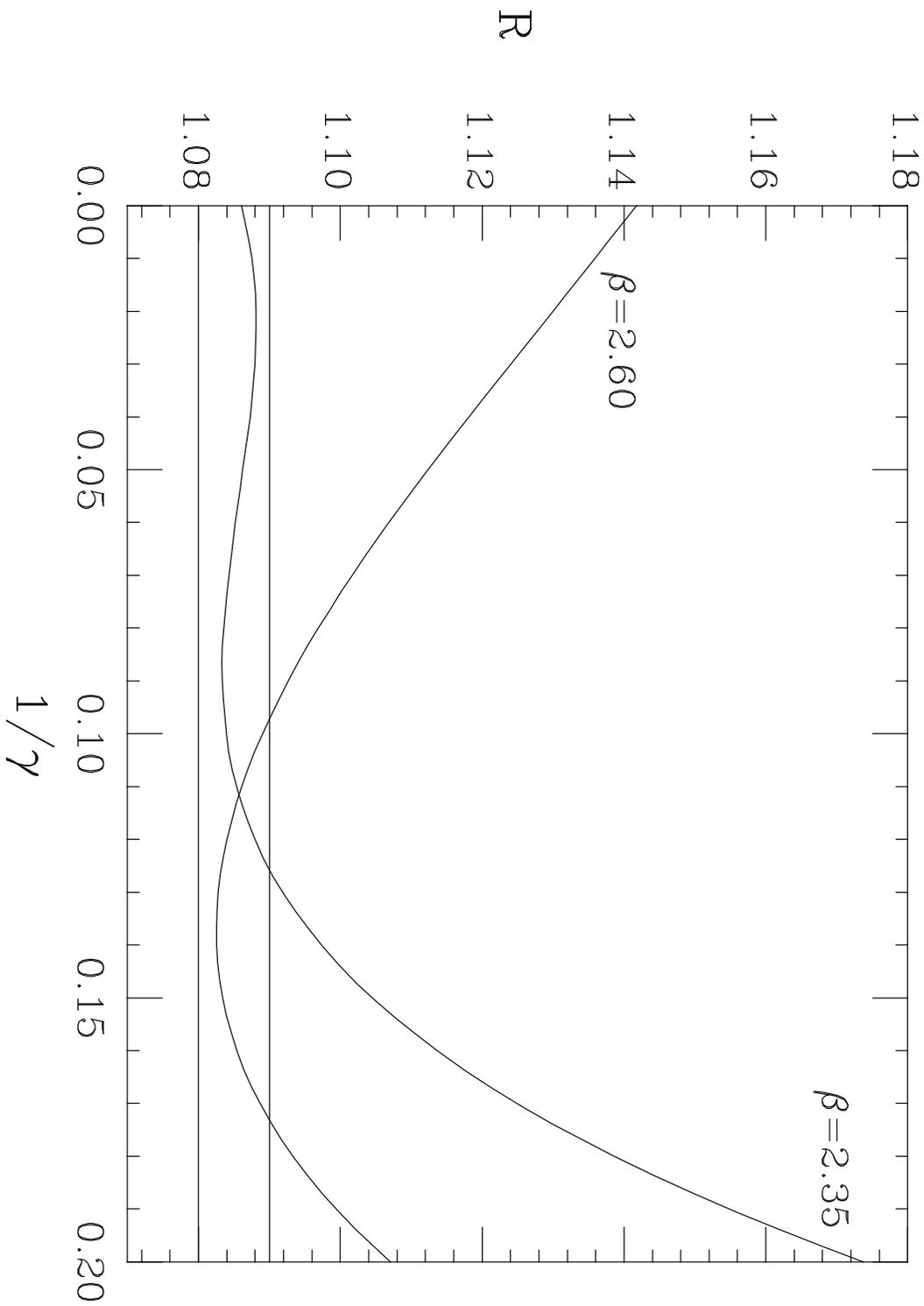
Fig. 3: $R(\beta,1/\gamma)$ for $\beta=2.35$ and $\beta=2.60$ vs. 
$1/\gamma$; the horizontal lines limit the scaling region.

\vskip 0.35 truecm
Fig. 4: Mass of the $\rho^+$ particle for $\beta=2.35$(a) and 

$\beta=2.60$(b) vs. $1/\gamma$.
\vskip 0.35 truecm

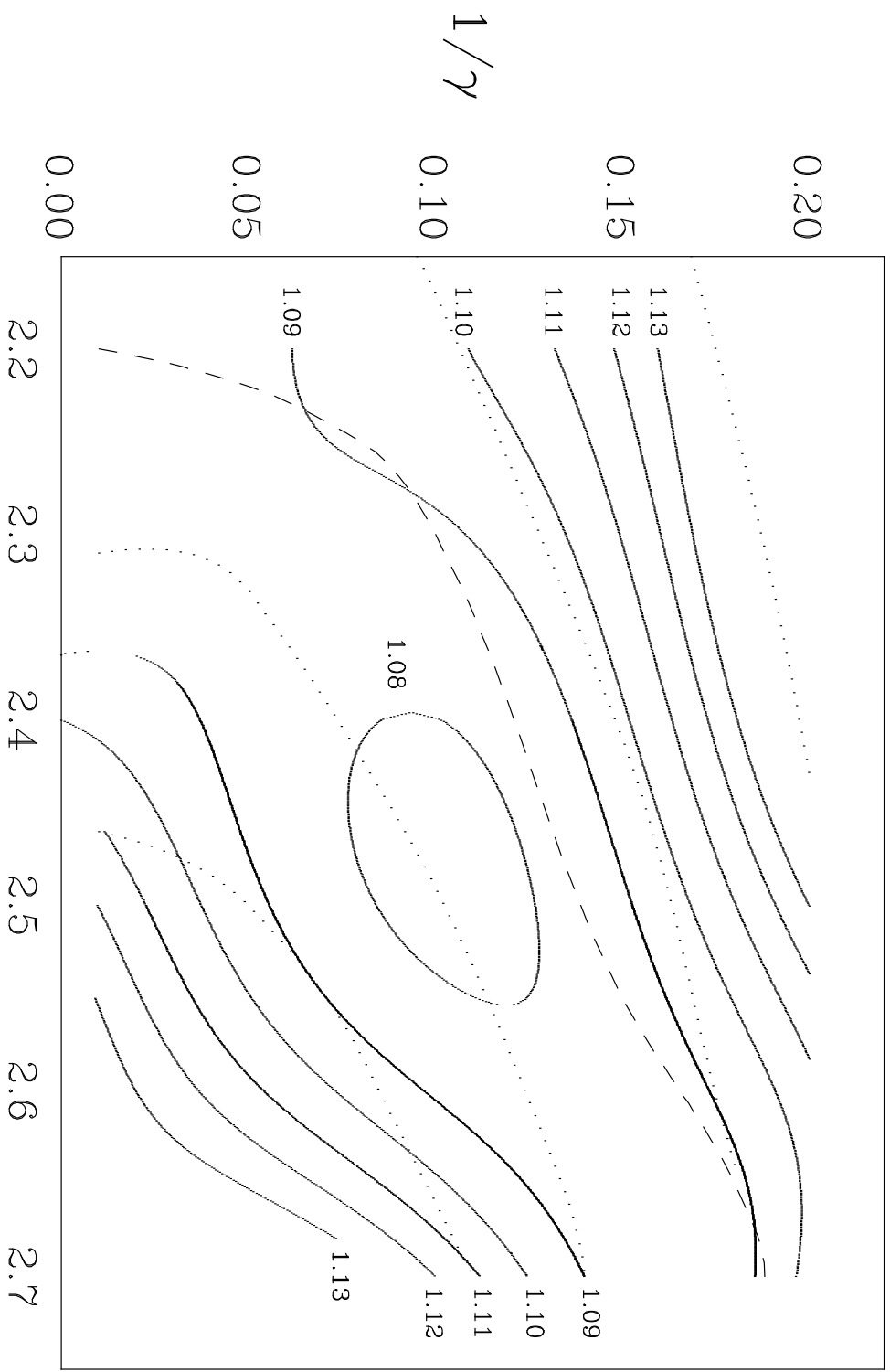
Fig. 5: Lines of constant $R$ (continuous), constant plaquette (dotted) and
position of the peak of specific heat (dashed) 

in the $\beta,1/\gamma$ plane.
\vskip 0.35 truecm

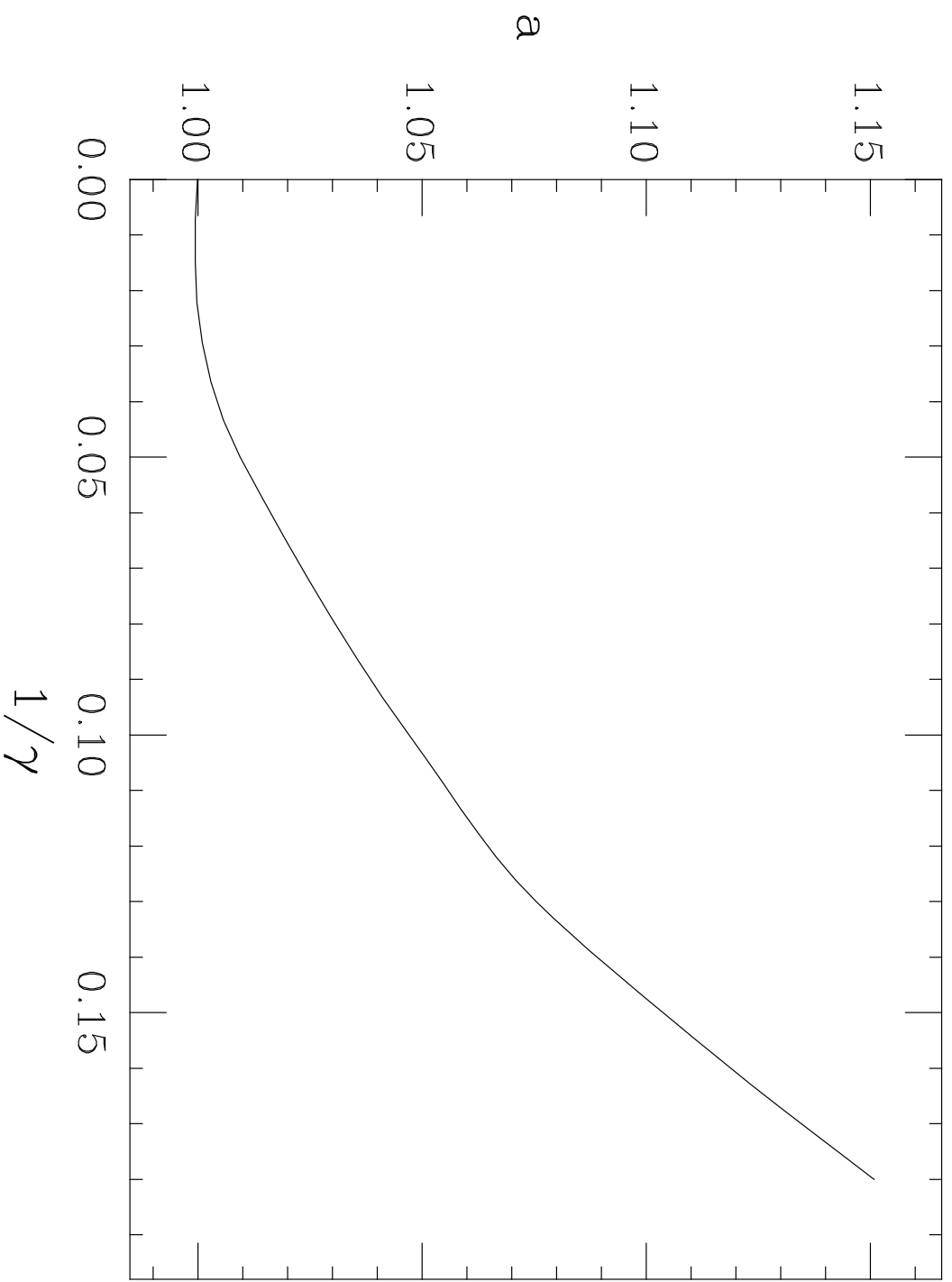
Fig. 6: Ratio of non-compact and Wilson lattice spacing along a
curve lying on the bottom of the scaling valley.
\vskip 0.35 truecm

\newpage
\pagestyle{empty}
\begin{figure}[!t]\
\psrotatefirst
\psfig{figure=fig1.epsi,angle=0,width=440pt}
\end{figure}

\newpage
\begin{figure}[!t]\
\psrotatefirst
\psfig{figure=fig2.epsi,angle=0,width=440pt}
\end{figure}

\newpage
\begin{figure}[!t]\
\psrotatefirst
\psfig{figure=fig3.epsi,angle=0,width=440pt}
\end{figure}

\newpage
\begin{figure}[!t]\
\psrotatefirst
\psfig{figure=fig4a.epsi,angle=0,width=440pt}
\end{figure}

\newpage
\begin{figure}[!t]\
\psrotatefirst
\psfig{figure=fig4b.epsi,angle=0,width=440pt}
\end{figure}

\newpage
\begin{figure}[!t]\
\psrotatefirst
\psfig{figure=fig5.epsi,angle=0,width=440pt}
\end{figure}

\newpage
\begin{figure}[!t]\
\psrotatefirst
\psfig{figure=fig6.epsi,angle=0,width=440pt}
\end{figure} 

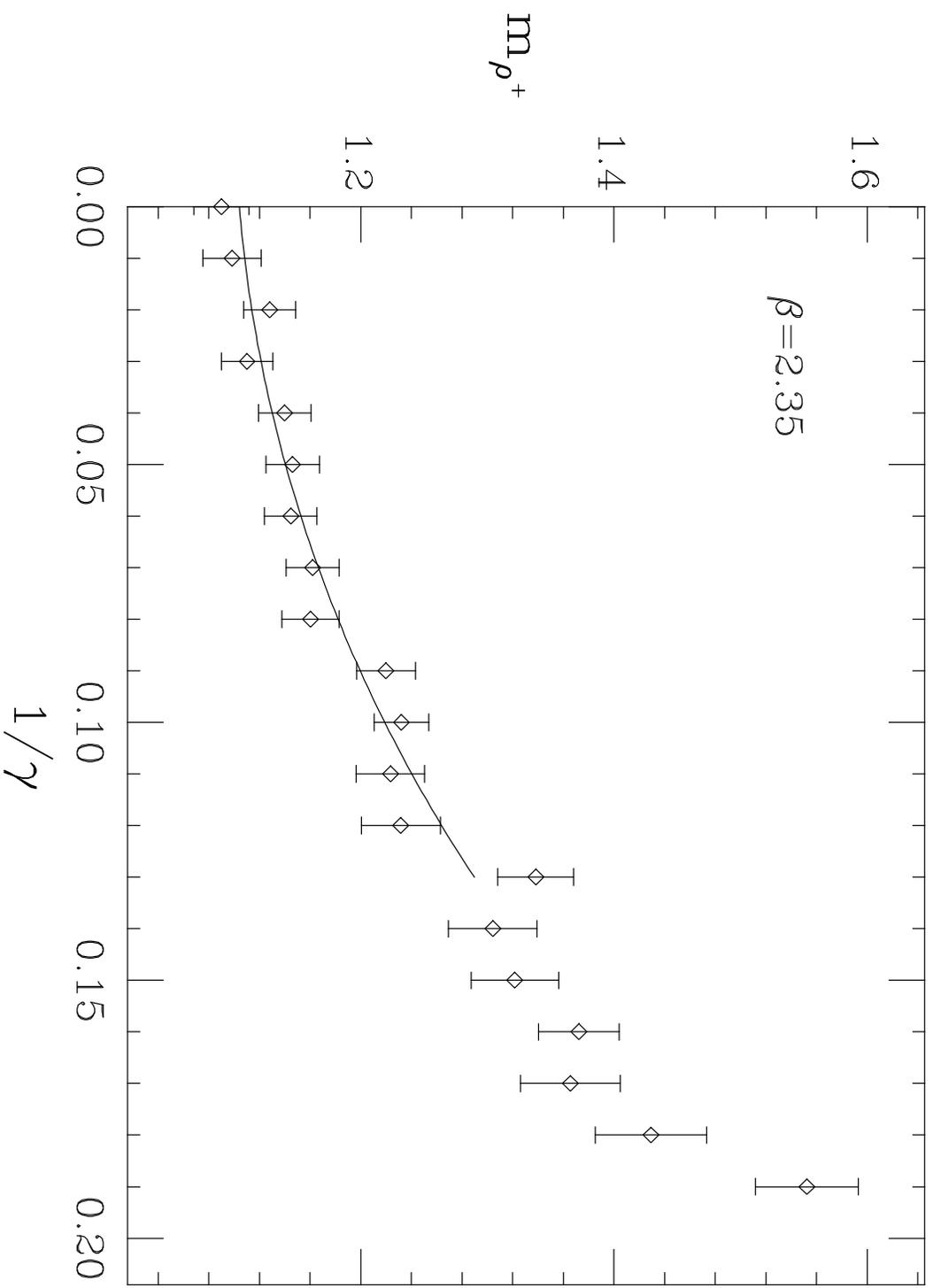
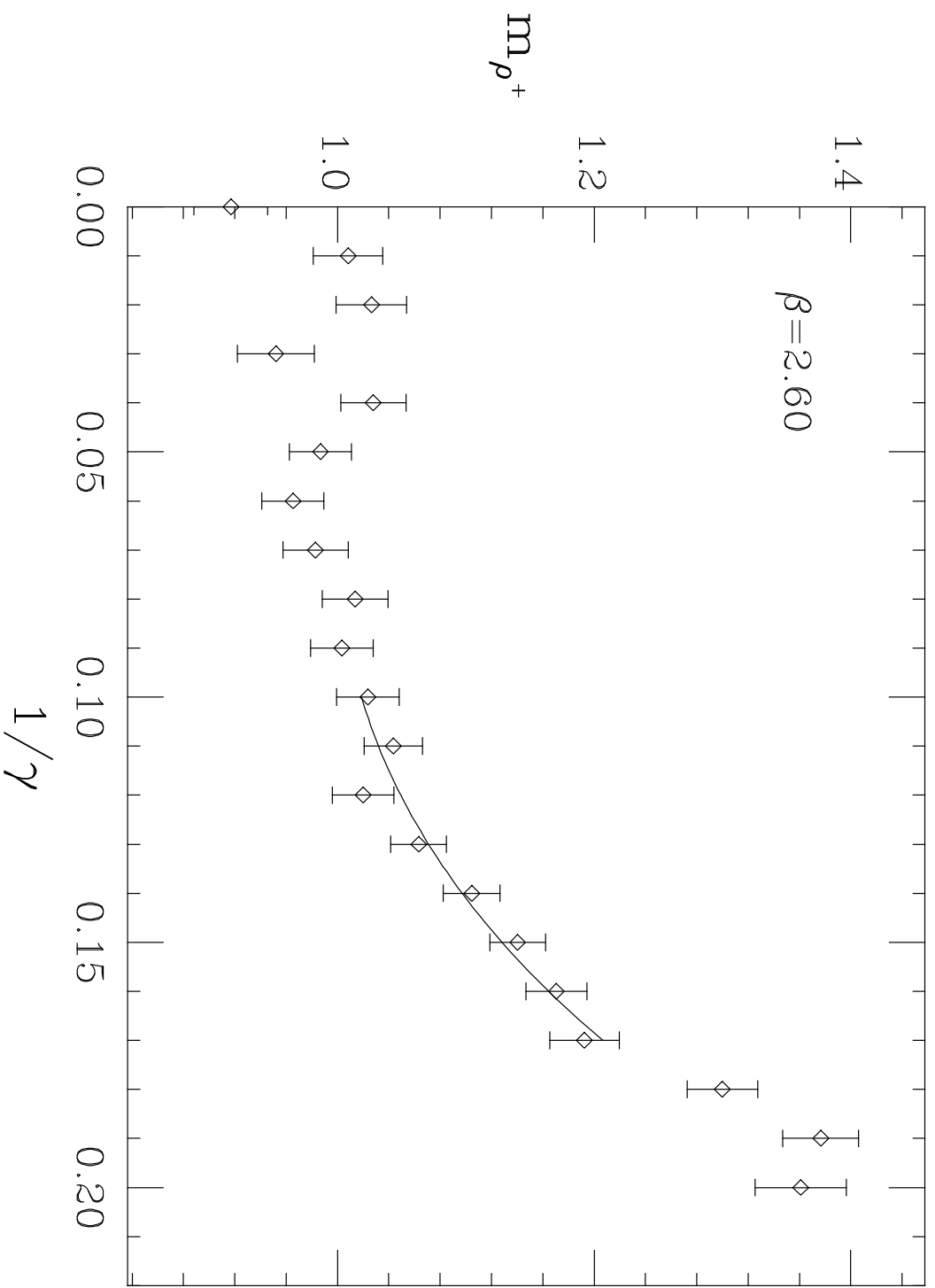
\end{document}